\documentclass[astrosymb]{aastex631}

\usepackage{newtxtext}
\usepackage{newtxmath}
\usepackage[T1]{fontenc}
\usepackage{graphicx}	

\usepackage{amsmath}	
\usepackage{amssymb}	
\usepackage{amsfonts}
\usepackage{xcolor}
\usepackage{multirow}
\usepackage{float}

\shorttitle{Model-Agnostic Cosmological Inference with SDSS-IV eBOSS}
\shortauthors{P. Mukherjee, A. A. Sen}

\graphicspath{{figs/}}

\begin{document}

\title{Model-Agnostic Cosmological Inference with SDSS-IV eBOSS: Simultaneous Probing for Background and Perturbed Universe}

\correspondingauthor{Purba Mukherjee} 

\author[0000-0002-2701-5654]{Purba Mukherjee}
\email{pdf.pmukherjee@jmi.ac.in}
\affiliation{Centre for Theoretical Physics, Jamia Millia Islamia, New Delhi-110025, India.}

\author[0000-0001-9615-4909]{Anjan A Sen}
\email{aasen@jmi.ac.in}
\affiliation{Centre for Theoretical Physics, Jamia Millia Islamia, New Delhi-110025, India.}



\begin{abstract}
Here we explore certain subtle features imprinted in data from the completed Sloan Digital Sky Survey IV (SDSS-IV) extended Baryon Oscillation Spectroscopic Survey (eBOSS) as a combined probe for the background and perturbed Universe. We reconstruct the baryon Acoustic Oscillation (BAO) and Redshift Space Distortion (RSD) observables as functions of redshift, using measurements from SDSS alone. We apply the Multi-Task Gaussian Process (MTGP) framework to model the interdependencies of cosmological observables $D_M(z)/r_d$, $D_H(z)/r_d$, and $f\sigma_8(z)$, and track their evolution across different redshifts. Subsequently, we obtain constrained three-dimensional phase space containing $D_M(z)/r_d$, $D_H(z)/r_d$, and $f\sigma_8(z)$ at different redshifts probed by the SDSS-IV eBOSS survey. Furthermore, assuming the $\Lambda$CDM model, we obtain constraints on model parameters $\Omega_{m}$, $H_{0}r_{d}$, $\sigma_{8}$ and $S_{8}$ at each redshift probed by SDSS-IV eBOSS. This indicates redshift-dependent trends in $H_0$, $\Omega_m$, $\sigma_8$ and $S_8$ in the $\Lambda$CDM model, suggesting a possible inconsistency in the $\Lambda$CDM model. Ours is a template for model independent extraction of information for both background and perturbed Universe using a single galaxy survey taking into account all the existing correlations between background and perturbed observables and this can be easily extended to future DESI-3YR as well as Euclid results.
\end{abstract}



\section{Introduction} \label{sec:intro}

The $\Lambda$CDM model has long been the cornerstone of modern cosmology, providing a robust framework to explain diverse phenomena \citep{SupernovaCosmologyProject:1998vns, SupernovaSearchTeam:1998fmf, Blanchard:2022xkk, Peebles:2024txt}, such as the temperature and polarization fluctuations of the cosmic microwave background (CMB) \citep{Planck:2015bue, Planck:2018vyg, ACT:2020gnv, Tristram:2023haj}, the large-scale structure of the Universe \citep{BOSS:2014hhw, BOSS:2016wmc, eBOSS:2020yzd, DESI:2024mwx}, and the distance-redshift relation of type Ia supernovae (SNIa) \citep{SDSS:2014iwm, Pan-STARRS1:2017jku, Brout:2022vxf, DES:2024jxu}. Despite its success, $\Lambda$CDM faces theoretical and observational challenges: theoretical concerns include the unresolved nature of dark matter \citep{Gaitskell:2004gd, LUX:2016ggv}, the cosmological constant and the cosmic coincidence problem \citep{Weinberg:1988cp, Sahni:1999gb, Carroll:2000fy, Padmanabhan:2002ji, Peebles:2002gy}. Observationally, tensions such as the $>5\sigma$ Hubble constant discrepancy ($H_0$) \citep{Hazra:2013dsx, Verde:2019ivm, Riess:2019qba, Riess:2021jrx, DiValentino:2020zio, Brieden:2022heh, Freedman:2023jcz, Efstathiou:2024dvn} between local distance ladder measurements and CMB-inferred values, as well as the $\sim2-2.5\sigma$ amplitude of matter fluctuations ($S_8$) \citep{DiValentino:2020vvd, Heymans:2020gsg, DES:2021wwk, Li:2023tui} tension between early CMB data and weak lensing surveys, remain unresolved. Recent observations by the James Webb Space Telescope have unveiled massive galaxies at unexpectedly high redshifts ($z \sim 15$) \citep{Labbe:2022ahb, Boylan-Kolchin:2022kae}, further challenging the concordance framework.  

Central to these investigations is understanding the energy composition of the Universe, the mechanisms driving cosmic expansion, and the growth of cosmic structures. To accomplish this, scientific models must deliver predictions that are both consistent with and relevant to these observations \citep{Bull:2015stt, DiValentino:2021izs, Abdalla:2022yfr, Perivolaropoulos:2021jda}. In cosmology, redshift $z$ acts as a proxy for time, making it vital to examine the $\Lambda$CDM parameters across redshift bins \citep{H0LiCOW:2019pvv, Millon:2019slk, Krishnan:2020obg, Krishnan:2020vaf, Krishnan:2022fzz, Dainotti:2021pqg, Colgain:2022nlb, Hu:2022kes, Jia:2022ycc, Colgain:2022rxy, Vagnozzi:2023nrq, Risaliti:2018reu, Lusso:2020pdb, Yang:2019vgk, Khadka:2020vlh, Pasten:2023rpc, Adil:2023jtu, Akarsu:2024hsu, Artis:2024zag, ACT:2024nrz}. For instance, trends of $H_0$ decreasing and $\Omega_m$ increasing, along with an increase of $\sigma_8$ and $S_8$ values from low to high $z$ reported in some recent studies, challenge the fundamental assumption of constancy of model parameters \citep{Krishnan:2020vaf, Krishnan:2022fzz}. These studies hint at possible missing physics at specific epochs, underscoring the importance of identifying redshift ranges where $\Lambda$CDM may break down. Such insights are essential for refining cosmological models and advancing our understanding of the Universe's evolution.

In this work, we analyze data exclusively from the completed Sloan Digital Sky Survey (SDSS)-IV extended Baryon Oscillation Spectroscopic Survey (eBOSS) \citep{eBOSS:2020yzd}, which has been instrumental in advancing cosmological analyses. The BOSS and eBOSS surveys have pioneered the use of Baryon Acoustic Oscillations (BAO) \citep{Eisenstein:1997ik} and Redshift Space Distortions (RSD) \citep{Guzzo:1997mq} to probe the Universe. Herein, we consider data from spectroscopic galaxy and quasar samples spanning four generations of SDSS, including SDSS MGS \citep{Howlett:2014opa}, BOSS galaxies \citep{BOSS:2016wmc}, eBOSS LRGs \citep{eBOSS:2020lta, eBOSS:2020hur}, eBOSS ELGs \citep{eBOSS:2020qek, eBOSS:2020fvk}, and eBOSS quasars \citep{eBOSS:2020gbb, eBOSS:2020uxp}, as well as Ly-$\alpha$ auto- and cross-correlation measurements from BOSS and eBOSS \citep{eBOSS:2020tmo}. By focusing solely on SDSS data, we avoid potential conflicts that may be present among datasets from disparate sources. This single-survey approach ensures that our results are less affected by inter-survey calibration errors, systematic uncertainties, modeling discrepancies and external biases that can complicate multi-survey analyses.  

Our analysis focuses on reconstructing BAO and RSD observables as a function of redshift. BAO features, observed in both transverse and line-of-sight directions, constrain cosmological distances, such as transverse comoving distance $D_M(z)/r_d$ and Hubble distance $D_H(z)/r_d$. Meanwhile, RSD effects \citep{Kaiser:1987qv}, caused by the bulk motion of matter in gravitational potential wells,  provide insights into structure formation through $f\sigma_8$, a parameter quantifying the peculiar velocity fields.  To this end, we employ the Multi-Task Gaussian Process (MTGP) \citep{Caruana1998, rw, NIPS2007_66368270}, a machine learning framework, to reconstruct the evolution of these BAO and RSD observables in a model independent manner as far as the late time cosmology is concerned. MTGP effectively models the complex interdependencies among $D_M(z)/r_d$, $D_H(z)/r_d$, and $f\sigma_8(z)$ measurements, while integrating systematic and statistical uncertainties directly into the covariance matrix. This also helps us to identify any possible presence of redshift-dependent trends in $H_0$, $\Omega_m$, and $S_8$ in $\Lambda$CDM.
\textbf{It also paves the way for future similar studies using the Dark Energy Spectroscopic Instrument (DESI) Full Shape measurements \citep{DESI:2024jis}, the direct successor to SDSS, serving as a promising diagnostic tool for upcoming analyses.}

This paper is organized as follows: Section \ref{sec:theory} outlines the key concepts and models that underpin the study. In section \ref{sec:method} the relevant data and reconstruction techniques used in the study are described. Section \ref{sec:result}  highlights the outcomes of the reconstruction process, followed by consistency checks for $\Lambda$CDM, and a comparison of our findings with complementary datasets to validate their robustness. Finally, we summarize the key insights and potential areas for future research in section \ref{sec:conclusion}.

\section{Theoretical Framework} \label{sec:theory}

On large scales, the Universe is described by the spatially flat, homogeneous, and isotropic Friedmann-Lema\^{i}tre-Robertson-Walker (FLRW) metric, which governs its background evolution. Within this framework, the \textit{Hubble distance}, 
\begin{equation}
    D_H(z) = \frac{c}{H(z)} \, ,
\end{equation} serves as a characteristic scale that relates the expansion rate of the Universe to distances, where $c$ is the speed of light and $H(z)$ is the Hubble parameter at redshift $z$. At the present epoch, this reduces to $D_H(z=0) = \frac{c}{H_0}$, where $H_0$ is the Hubble constant. Additionally, the \textit{comoving distance}, $D_M$, quantifies the separation between two points in the Universe while accounting for its expansion. For a source at redshift $z$, it is defined as
\begin{equation}
D_M(z) = c \int_0^z \frac{\mathrm{d}z'}{H(z')} \, .    
\end{equation}
These distances provide a foundation for interpreting cosmological observations and understanding the large-scale structure of the Universe. 

The Hubble parameter $H(z)$, which dictates the rate of expansion, is dependent on the underlying cosmological model. In the standard $\Lambda$-cold dark matter ($\Lambda$CDM) framework, for instance, it is given by
\begin{equation}
H(z) = H_0 \sqrt{\Omega_m(1+z)^3 + \Omega_\Lambda} \, ,
\end{equation}
where $\Omega_m$ and $\Omega_\Lambda$ are the present matter and dark energy density parameters, respectively. This model assumes a spatially flat Universe, with matter and dark energy (described by a cosmological constant) as the primary components driving the evolution of the cosmos. Other cosmological models, such as those that incorporate dynamical dark energy behavior or modifications to gravity, can lead to different functional forms for $H(z)$. In these models, the Hubble parameter could be influenced by parameters such as the equation of state of dark energy, $w(z)$, or modifications to the Friedmann equations that account for the effects of new physics on the expansion rate \citep{DiValentino:2021izs, Abdalla:2022yfr}. Thus, the form of $H(z)$ is a key signature of the cosmological model in question and plays a crucial role in interpreting observational data. 

The evolution of cosmic structures is governed by the dynamics encoded in the Hubble parameter, which directly impacts the growth rate of perturbations. At the perturbation level, the growth rate of cosmic structures provides a key insight into the evolution of matter density fluctuations and the underlying cosmological model. It is commonly expressed through the observable $f \sigma_8$, which combines the linear growth rate of structures, $f$, with $\sigma_8$, the root-mean-square (rms) fluctuation of the matter density field in spheres of radius $8 \, h^{-1} \, \mathrm{Mpc}$.

The growth rate $f$ is defined as,
\begin{equation}
f = \frac{\mathrm{d} \ln D(a)}{\mathrm{d} \ln a}  = -(1+z)\frac{D^\prime(z)}{D(z)} \, ,     
\end{equation}
where $D(a)$ is the linear growth factor. Under general relativity, $f$ can often be approximated as $f \approx \Omega_m(a)^\gamma$, where $\Omega_m(a)$ is the matter density parameter at scale factor $a$, and $\gamma$ is the growth index, typically around $\gamma \approx 0.55$. The parameter $\sigma_8$ quantifies the amplitude of matter density fluctuations and is influenced by the normalization of the initial power spectrum, defined as
\begin{equation}
    \sigma_8(z) = \sigma_8 (z=0) D(z) \, .
\end{equation}
Thus, the product $f \sigma_8$ serves as a valuable probe, combining information on the rate of structure formation and the amplitude of clustering.

\section{Data and Methodology} \label{sec:method}

The BAO observations probe the large-scale structure of the Universe, providing insights into its geometry and the growth of structures. These observations are quantified using normalized distances relative to the sound horizon at the baryon drag epoch, denoted $r_d$, which is the distance sound waves travelled from the Big Bang to the epoch of baryon drag \citep{Eisenstein:1997ik}, defined as
\begin{equation}
r_d = \int_{z_d}^{\infty} \frac{c_s(z)}{H(z)} \mathrm{d}z \, ,
\end{equation}
where $z_d$ is the redshift of the drag epoch and $c_s$ is the sound speed.

In spectroscopic surveys, the BAO feature appears along both the line of sight and the transverse direction. Along the line of sight, the redshift interval $\Delta z$ directly measures the Hubble parameter $H(z) = \frac{c \Delta z}{r_d}$, with the Hubble distance $D_H(z)$. In the transverse direction, the BAO scale corresponds to an angular separation $\Delta \theta$, enabling the estimation of the comoving angular diameter distance $D_M(z) = \frac{r_d}{\Delta \theta}$. Galaxy redshift measurements from spectroscopic BAO surveys also reveal anisotropic clustering, influenced by the Redshift Space Distortion (RSD) \citep{Kaiser:1987qv, Guzzo:1997mq}. The RSD effect, driven by the growth of structure and peculiar velocities, introduces additional redshifts along the line of sight, leading to anisotropic clustering, which is tied to the growth rate $f \sigma_8$. Together, BAO and RSD measurements, $\frac{D_M(z)}{r_d}$, $\frac{D_H(z)}{r_d}$, and $f\sigma_8(z)$, can provide robust constraints on the expansion history and structure growth of the Universe. 

In this study, we employ the Multi-Task Gaussian Process (MTGP) \citep{Haridasu:2018gqm, Perenon:2021uom, Mukherjee:2024ryz, Dinda:2024ktd}, a machine learning technique, to analyze the evolution of BAO and RSD observables across multiple generations of SDSS data, spanning a redshift range of $0 < z < 2.34$. The data compilation includes measurements from various tracers in different redshift intervals, summarized in Table \ref{tab:data}. Unlike traditional approaches that combine these observations with external datasets, viz. Planck \citep{Planck:2018vyg, Tristram:2023haj} or Type Ia supernovae \citep{Brout:2022vxf, DES:2024jxu}, or the informed use of cosmological priors \citep{Peirone:2017lgi, Patel:2024odo, Payeur:2024dnq}, we focus on directly extracting features from the BAO and RSD measurements within the SDSS data alone, ensuring a systematics-minimized analysis.

While a single-task GP (see \citet{PhysRevLett.105.241302, PhysRevD.84.083501, Seikel:2012uu, Shafieloo:2012ht, Mukherjee:2022lkt, Ghosh:2024kyd} and references therein), is effective for reconstructing individual functions from independent datasets, it does not account for the shared information between observables derived from overlapping datasets. In our case, the observables $D_M(z)/r_d$, $D_H(z)/r_d$, and $f\sigma_8(z)$, derived from galaxies, quasars, and Ly-$\alpha$ forests across different redshift ranges, exhibit interdependencies and are influenced by common systematics and statistical uncertainties, governed by the same underlying physics. Treating each observable independently risks underestimating uncertainties and leading to suboptimal reconstructions.

The MTGP framework overcomes this limitation by modeling redshift-dependent relationships between the observables - their auto-correlations and cross-correlations through a joint covariance structure. We use three squared exponential kernels to model the individual functions, 
\begin{equation}
    k_{i \times i}(z, \tilde{z}) = {\sigma_f}^2_i \exp \left[ - \frac{\left(z - \tilde{z}\right)^2}{2 l_i^2} \right] \, \hspace{3.5cm} \cdots~ i = 1,~2,~3 \,
\end{equation}
and a convolution of two kernels 
\begin{equation}
    k_{i \times j}(z, \tilde{z}) = {\sigma_f}_i {\sigma_f}_j \left(\frac{2 l_i l_j}{l_i^2 + l_j^2}\right)^{\frac{1}{2}} \exp \left[ - \frac{\left(z - \tilde{z}\right)^2}{l_i^2 + l_j^2} \right]  \, \hspace{1.3cm} \cdots~ i = 1,~2,~3 \,
\end{equation}
to capture the correlations between them. Here, $\left\lbrace {\sigma_f}_i, l_i\, , \cdots \, i = 1,2,3 \right\rbrace$ are the hyperparameters of the kernel, which are trained by marginalizing over the log-likelihood, 
\begin{equation}
    \ln \mathcal{L} \left(\left\lbrace {\sigma_f}_i, l_i\, , \cdots \, i = 1,2,3 \right\rbrace \right) = -\frac{1}{2} y^\text{T} \left( \tilde{K} + \mathcal{C} \right)^{-1} y - \frac{1}{2} \ln \vert \tilde{K} + \mathcal{C} \vert - \frac{n}{2} \ln 2\pi \, .
\end{equation}
Here, $n$ is the total number of SDSS data points, $\tilde{K} = \left[ K_{ij} \right]$ is the joint MTGP kernel, $y = \left[\frac{D_M}{r_d} ~~ \frac{D_H}{r_d} ~~ f\sigma_8 \right]^{\text{T}} $ is the data, and
$$\mathcal{C} = \left[ \begin{matrix}
\text{cov}\left(D_M/r_d, \, D_M/r_d\right) &  \text{cov}\left(D_M/r_d, \, D_H/r_d\right) & \text{cov}\left(D_M/r_d, \, f\sigma_8\right) \\
\text{cov}\left(D_H/r_d, \, D_M/r_d\right) &  \text{cov}\left(D_H/r_d, \, D_H/r_d\right) & \text{cov}\left(D_H/r_d, \, f\sigma_8\right) \\
\text{cov}\left(f\sigma_8, \, D_M/r_d\right) &  \text{cov}\left(f\sigma_8, \, D_H/r_d\right) & \text{cov}\left(f\sigma_8, \, f\sigma_8\right) \, ,\\
\end{matrix} \right] $$ is the combined data covariance in block form. Finally, the predicted mean and covariance are,
\begin{align}
    \overline{f^\star} &= \tilde{K}^{\star} {\left[\tilde{K} + \mathcal{C} \right]}^{-1} y \,  \\
    \text{cov}{f^\star} &= \tilde{K}^{\star \, \star} - \tilde{K}^{\star} {\left[\tilde{K} + \mathcal{C} \right]}^{-1} \tilde{K}^{\star \, \text{T}} \, .
\end{align}
Therefore, our approach ensures a cohesive reconstruction by accounting for the interdependencies between the tracers and properly incorporating both systematic errors and statistical uncertainties into the covariance matrix. By leveraging these correlations, the MTGP framework enables a more accurate reconstruction of the redshift-dependent trends in the observables.

\begin{deluxetable*}{lcccccc}
\setcounter{table}{0}
\tablecaption{Summary of BAO and RSD observables for various tracers in SDSS-IV Data \label{tab:data}}
\tablewidth{\textwidth}
\tablehead{
\colhead{Tracer} & \colhead{$z_\text{eff}$} & \colhead{$D_M(z)/r_d$} & \colhead{$D_H(z)/r_d$} & \colhead{$f\sigma_8(z)$} & \colhead{Reference}
}
\startdata
MGS                  & 0.15   & \nodata & \nodata & $0.53 \pm 0.16$  & \citet{Howlett:2014opa}\\
BOSS Galaxy (low-$z$)  & 0.38   & $10.27 \pm 0.15$ & $24.89 \pm 0.58$ & $0.497 \pm 0.045$  & \citet{BOSS:2016wmc} \\
BOSS Galaxy (high-$z$) & 0.51   & $13.38 \pm 0.18$ & $22.43 \pm 0.48$ & $0.459 \pm 0.038$  & \citet{BOSS:2016wmc} \\
eBOSS LRG            & 0.698  & $17.65 \pm 0.30$ & $19.78 \pm 0.46$ & $0.473 \pm 0.041$  & \citet{eBOSS:2020yzd} \\
eBOSS QSO            & 1.48   & $30.21 \pm 0.79$ & $13.23 \pm 0.47$ & $0.462 \pm 0.045$  & \citet{eBOSS:2020gbb} \\
Ly$\alpha$ QSO       & 2.334 & $37.5 \pm 1.2$  & $8.99 \pm 0.19$  & \nodata  & \citet{eBOSS:2020tmo} \\
\enddata
\end{deluxetable*}


\begin{figure*}[ht!]
	\centering
	\begin{minipage}{0.48\textwidth}
		\centering
		\includegraphics[width=\textwidth]{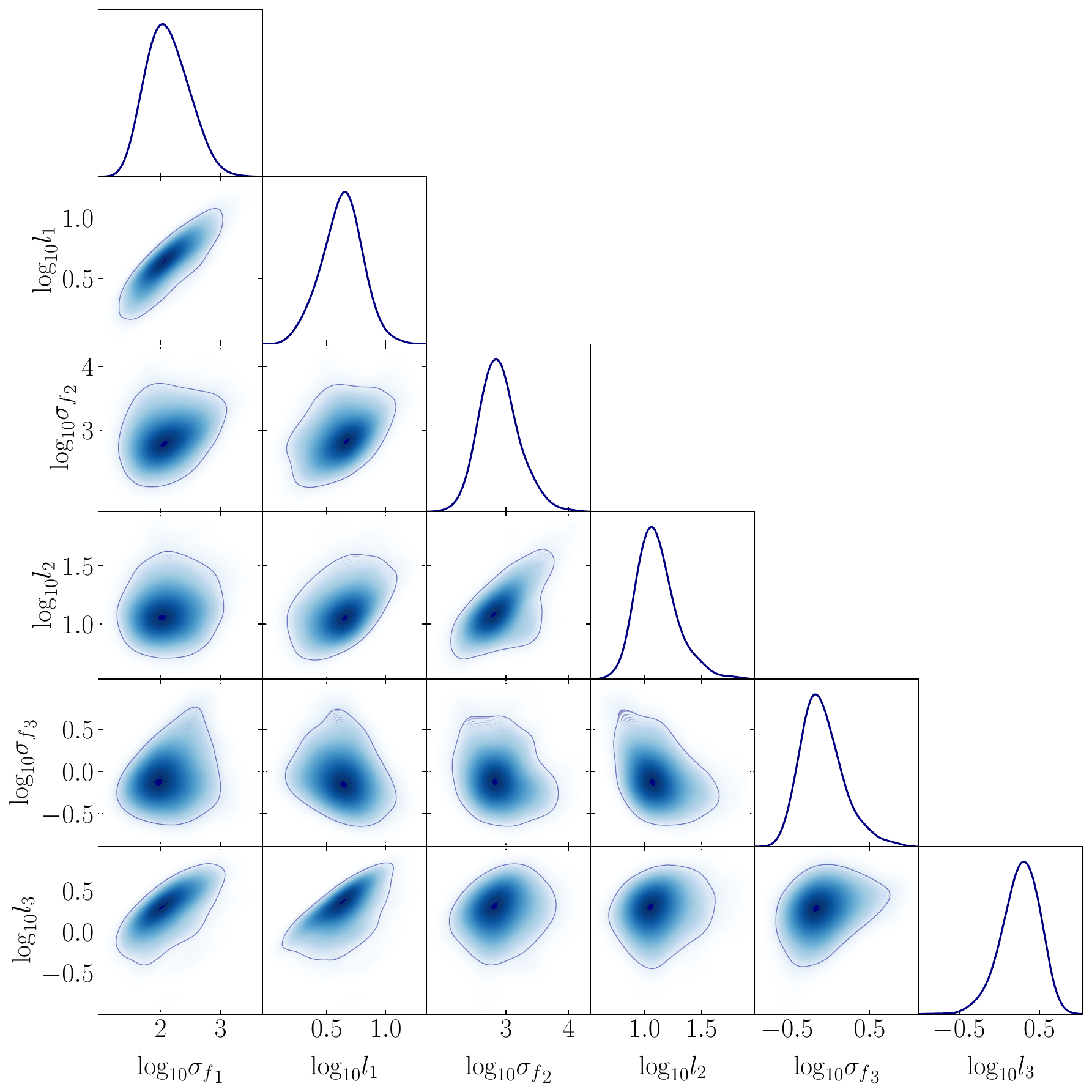}
		\caption{Triangle plot for MTGP hyperparameter samples.}
		\label{fig:gp}
	\end{minipage}%
	\hfill
	\begin{minipage}{0.48\textwidth}
		\centering
		{\bf Table 2.}~Hyperparameter values with best-fit and mean with $1\sigma$. \\[2pt]
		\begin{ruledtabular}
			\begin{tabular}{l c c c}
				Hyperparameters & Priors & Best-Fit & Mean with 1$\sigma$ \\
				\colrule
				${\log_{10}}{\sigma_f}_1$  & $\mathcal{U}[-5,5]$  & ~~$2.115$ & ~~$2.088_{-0.337}^{+0.411}$   \\
				${\log_{10}}{l_1}$ & $\mathcal{U}[-5,5]$  & ~~$0.629$ & ~~$0.640_{-0.187}^{+0.157}$   \\
				${\log_{10}}{\sigma_f}_2$  & $\mathcal{U}[-5,5]$  &  ~~$2.891$  & ~~$2.869_{-0.291}^{+0.336}$  \\
				${\log_{10}}{l_2}$  & $\mathcal{U}[-5,5]$  & ~~$1.107$ & ~~$1.085_{-0.154}^{+0.189}$  \\
				${\log_{10}}{\sigma_f}_3$ & $\mathcal{U}[-5,5]$ &  $-0.075$ & $-0.104_{-0.218}^{+0.273}$ \\
				${\log_{10}}{l_3}$ & $\mathcal{U}[-5,5]$   & ~~$0.265$ & ~~$0.281_{-0.255}^{+0.229}$  \\
			\end{tabular}
		\end{ruledtabular}
		\label{tab:gp}
	\end{minipage}
\end{figure*}

\section{Analysis and Discussions} \label{sec:result}

We undertake MTGP regression on the joint SDSS BAO+RSD data using the \texttt{tinygp}\footnote{\url{https://github.com/dfm/tinygp.git}} \citep{tinygp} module, implementing a Bayesian MCMC analysis with \texttt{jax}\footnote{\url{https://github.com/jax-ml/jax.git}} \citep{jax2018github} and \texttt{numpyro}\footnote{\url{https://github.com/pyro-ppl/numpyro.git}} \citep{numpyro, bingham2019pyro}. For this, we assume uniform flat priors on the kernel hyperparameters, as detailed in Table \ref{tab:gp}. The signal amplitudes $\log_{10} {\sigma_f}_i$ and length scales ${\log_{10} l}_i$ for each observable are optimized within the prior range of $[-5, 5]$. Large values of $\sigma_f$ for the two BAO observables $D_M/r_d$ and $D_H/r_d$ indicate strong signal strengths, leading to substantial contributions from these components to the overall covariance. In contrast, a lower $\sigma_f$ value for the RSD observable $f\sigma_8$ implies relatively lower variability or weaker correlations, which could stem from the smaller effective sample size or increased uncertainties associated with $f\sigma_8$ data. The length scales $l$ exhibit moderate values across all observables, suggesting a balance between the smoothness of the kernel and the flexibility to adapt to redshift-dependent variations in the datasets. The 1$\sigma$ uncertainties around the mean hyperparameter values are relatively small, indicating that the posterior distributions are well-constrained and that the data provide robust constraints on the kernel parameters. The marginalized posterior distributions and the corresponding 2D parameter spaces for the samples are visualized in Fig. \ref{fig:gp}, generated with \texttt{GetDist}\footnote{\url{https://github.com/cmbant/getdist.git}} \citep{Lewis:2019xzd}.

\subsection{Result of Reconstruction}

\begin{figure*}
    \centering
    \includegraphics[width=\linewidth]{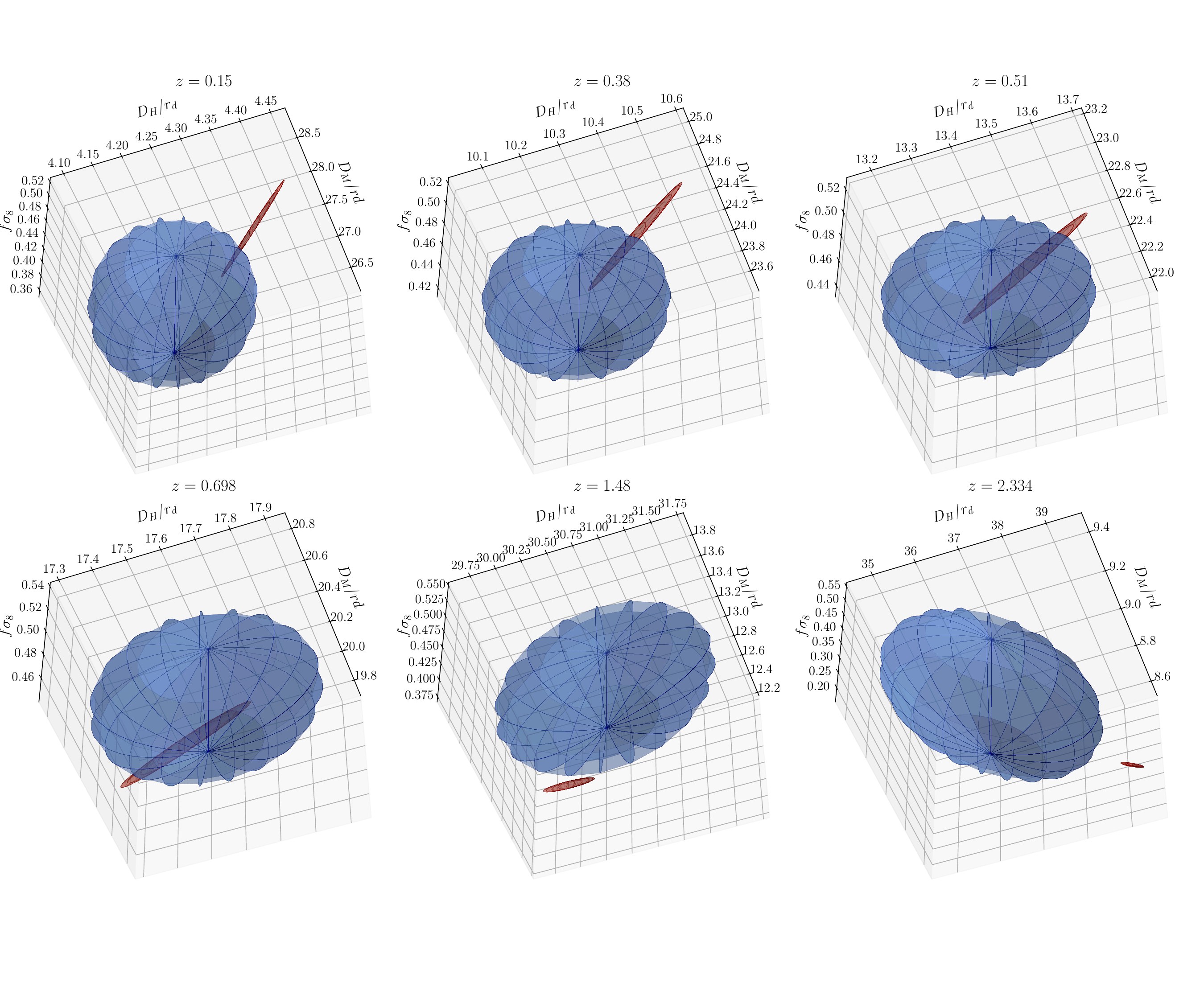} 
    \caption{3D phase spaces from the reconstructed functions $D_M/r_d$, $D_H/r_d$ and $f\sigma_8$ covering  $2\sigma$ uncertainty at SDSS effective redshifts.}
    \label{fig:rec_spheres}
\end{figure*}

Fig. \ref{fig:rec_spheres} displays six 3D phase spaces, corresponding to six redshift values, showcasing the reconstructed observables $D_M(z)/r_d$, $D_H(z)/r_d$, and $f\sigma_8(z)$ at the $2\sigma$ confidence level, obtained using the MTGP framework applied to SDSS BAO and RSD data. These plots offer a comprehensive visualization of the interplay between the background and perturbation sectors of cosmology. The MTGP reconstructions are shown in blue regions, whereas Planck $\Lambda$ CDM predictions are represented in red regions, allowing a direct comparison of their behavior across different redshifts. Each phase portrait captures the relationships between the three predicted observables at a specific redshift, providing a geometric perspective on their mutual correlations within parameter space. Consistent overlap between the blue and red regions indicates agreement between the MTGP reconstruction and $\Lambda$CDM, while noticeable deviations while deviations in specific observables may highlight potential tensions or the presence of new physics. For instance, 
\begin{itemize}
    \item At lower redshifts, $z=0.15, ~ 0.38, ~ 0.51$ and $0.698$ the phase spaces exhibit good agreement between the reconstructions and $\Lambda$CDM predictions.
    \item At higher redshifts $z=1.48$ and $z=2.334$ noticeable discrepancies from Planck $\Lambda$CDM are seen to emerge. 
\end{itemize} 
Therefore, these phase spaces are instrumental in analyzing the interplay between the background observables ($D_M(z)/r_d$ and $D_H(z)/r_d$) and the perturbation observable ($f\sigma_8(z)$), allowing for a joint assessment of the concordance model's performance across redshifts. It helps identify where and how deviations arise, offering insights into potential breakdowns of $\Lambda$CDM. It also highlights specific redshifts where the exploration of new physics could be motivated, providing a framework to explain the features observed in the data and guiding investigations beyond $\Lambda$CDM. 

\begin{figure*}
    \centering
    \includegraphics[width=\linewidth]{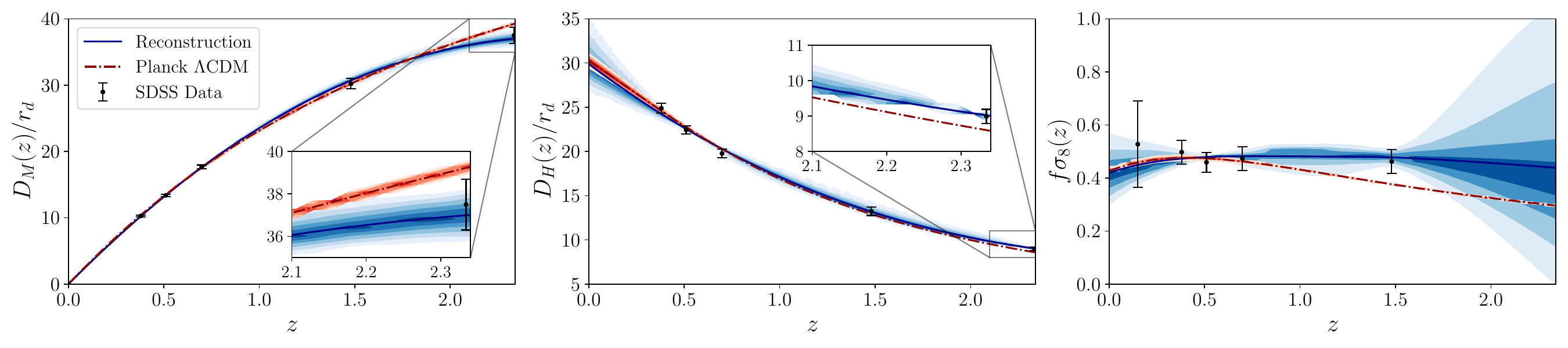} 
    \caption{Plots for the reconstructed functions $D_M(z)/r_d$, $D_H(z)/r_d$ and $f\sigma_8(z)$ [best-fit results with 1$\sigma$ \& 2$\sigma$ uncertainties] vs redshift in blue. Planck $\Lambda$CDM predictions are in red.}
    \label{fig:recz}
\end{figure*}

To better understand the observables when deviations from $\Lambda$CDM arise, we plot the reconstructed redshift evolution of the cosmological observables $D_M(z)/r_d$, $D_H(z)/r_d$, and $f\sigma_8(z)$ in Fig. \ref{fig:recz}. The best-fit lines from MTGP predictions are shown in blue with shaded confidence intervals plot using \texttt{fgivenx}\footnote{\url{https://github.com/handley-lab/fgivenx.git}} \citep{fgivenx}. The black points with error bars represent the observational data, while the red lines (along with the shaded confidence regions) show the corresponding predictions from the Planck 2018 $\Lambda$CDM model for comparison, offering insights into the consistency and potential tensions between the data and the standard cosmological model. The MTGP reconstruction closely aligns with the data, with well-constrained confidence intervals capturing the uncertainties. By extrapolating the reconstructed $D_H(z)/r_d$ to $z=0$, we obtain a constraint of $D_H/r_d(z=0) = 29.825 \pm 0.826$. This leads to a model-independent measurement of $H_0 r_d = 100.59 \pm 2.78$ in units of 100 km/s. Using the sound horizon $r_d$ inferred from early-Universe observations (which is completely independent of physics at low redshifts) as obtained by Planck $r_d = 147.09 \pm 0.26$ Mpc \citep{Planck:2018vyg}, we derive an inferred value of $H_0 = 68.38 \pm 1.89$ km Mpc$^{-1}$ s$^{-1}$. This result lies within 2$\sigma$ of both the Planck $\Lambda$CDM determination ($H_0^{\rm P18} = 67.36 \pm 0.54$ km Mpc$^{-1}$ s$^{-1}$ \citet{Planck:2018vyg}) and the SH0ES local measurement ($H_0^{\rm SH0ES} = 73.2 \pm 1.04$ km Mpc$^{-1}$ s$^{-1}$ \citet{Riess:2021jrx}). The error in the measured value of $H_{0}$ is around $2.76\%$ from SDSS-IV eBOSS solely, given an early Universe Prior. This shows the potential of MTGP framework in determining cosmological parameters in a model-independent way from a single survey like SDSS-IV. 

\noindent Furthermore, we notice the following trends:

\begin{itemize}
    \item \textbf{Agreement with Planck $\Lambda$CDM in lower redshifts}: The reconstructed trends for background observables $D_M(z)/r_d$ and $D_H(z)/r_d$, derived purely from the background expansion history, are consistent with Planck predictions in the redshift range $z<1.48$ at the 1$\sigma$ confidence level. The reconstructed $f\sigma_8(z)$, which probes the growth of linear perturbations, also agrees with Planck $\Lambda$CDM in the redshift range $z \lesssim 1$. 
    \item \textbf{Deviations in $f\sigma_8(z)$ at higher redshifts}: A statistically significant deviation exceeding $2\sigma$ arises in $f\sigma_8$, suggesting possible tensions with Planck $\Lambda$CDM and hinting at potential new physics affecting the perturbation sector. No such deviations are found in $D_M(z)/r_d$ and $D_H(z)/r_d$, which remain consistent with the Planck baseline model. 
    \item \textbf{Anomalies in $D_M(z)/r_d$ and $D_H(z)/r_d$ at higher redshifts}: An additional deviation at $z = 2.334$ is observed in the $D_M(z)/r_d$ and $D_H(z)/r_d$ reconstruction relative to the Planck predictions. This feature is difficult to interpret due to the absence of corresponding $f\sigma_8(z)$ measurements, leaving it unclear whether it signifies new physics or a statistical anomaly. 
\end{itemize}
\noindent These trends highlight the standard cosmological model's consistency at lower redshifts, emphasizing the need for further investigation into the deviations at higher redshifts. The significant $2\sigma$ deviation at $z=1.48$ in $f\sigma_8(z)$ strongly points to potential tensions with $\Lambda$CDM, while the $z=2.334$ point may reflect as an outlier in the background sector. Determining whether these anomalies arise from unmodeled systematics, statistical fluctuations, or indications of beyond-$\Lambda$CDM physics requires additional scrutiny. These findings underscore the need for combining data from both the background and perturbation sectors to fully understand deviations from the standard cosmological framework.

\subsection{Consistency checks for $\Lambda$CDM}

\begin{figure*}
    \centering
    \includegraphics[width=\linewidth]{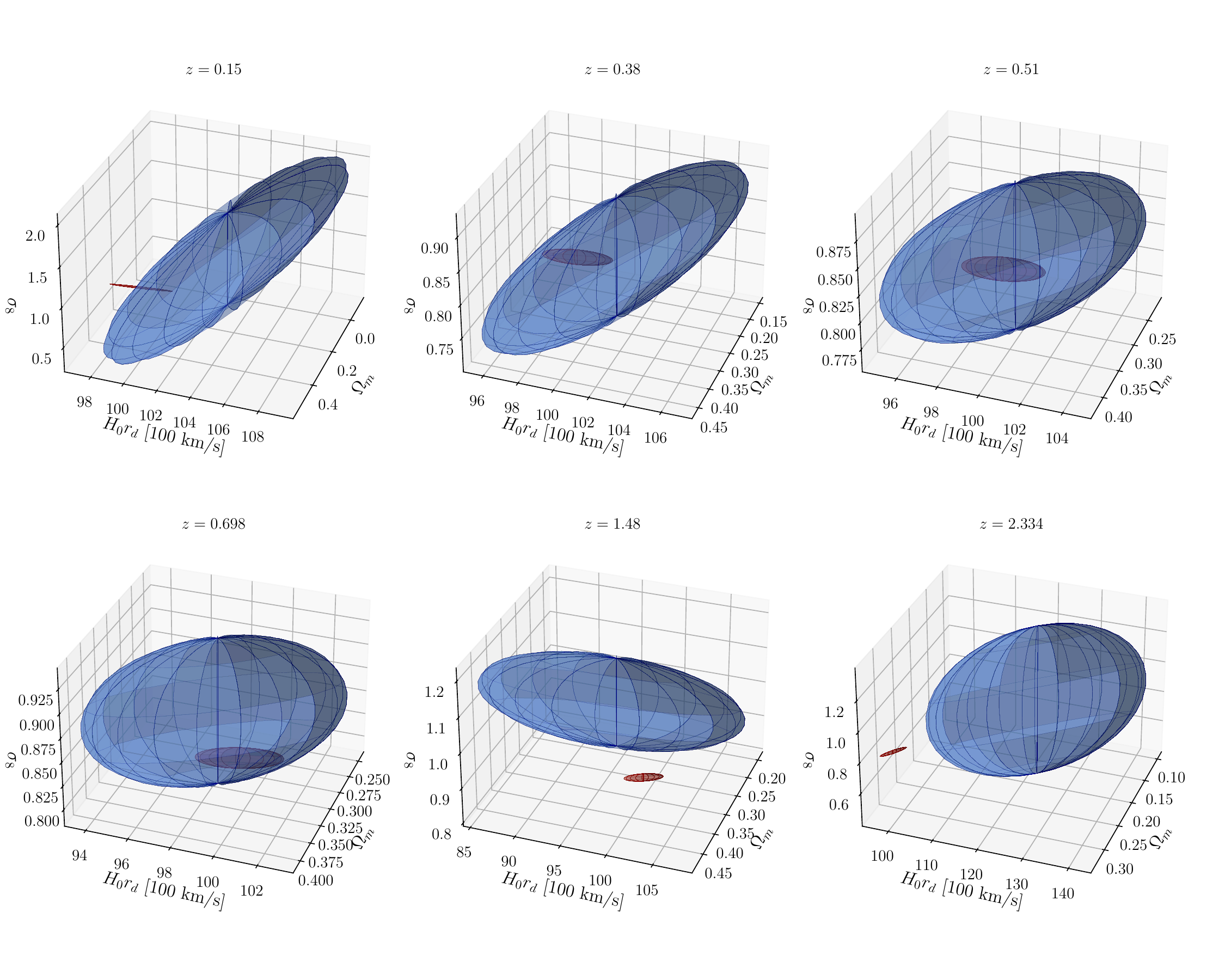} 
    \caption{3D phase spaces of the $\Lambda$CDM parameters (best-fit with $2\sigma$ uncertainty) obtained on fitting the $\Lambda$CDM model to the reconstructed functions $D_M/r_d$, $D_H/r_d$ and $f\sigma_8$ at SDSS effective redshifts in blue. Planck $\Lambda$CDM predictions are in red. }
    \label{fig:lcdm_spheres}
\end{figure*}

In what follows, we fit the parameters of the $\Lambda$CDM model to the individual 3D reconstructed phase spaces of the observables $D_M(z)/r_d$, $D_H(z)/r_d$, and $f\sigma_8(z)$ for each redshift bin, using \texttt{Cobaya}\footnote{\url{https://github.com/CobayaSampler/cobaya.git}} \citep{Torrado:2020dgo}. Table \ref{tab:lcdm} summarizes the resulting parameter estimates, including the best-fit values and the means with their 1\(\sigma\) uncertainties for $H_0 r_d$, $\Omega_m$, $\sigma_8$, and $S_8$. These fits provide a detailed assessment of how the reconstructed datasets at each effective redshift $z_{\rm eff}$, their mutual correlations, comply with the predictions of $\Lambda$CDM. It captures how effectively the baseline model explains the intricate interplay between the background and perturbation sectors, offering valuable insights into the trends and potential deviations of the cosmological parameters across redshift as a consistency check for the underlying model. 

For completeness, we also plot the 3D phase spaces of the reconstructed $\Lambda$CDM parameters ($H_0 r_d$, $\Omega_m$, and $\sigma_8$) for six distinct redshift bins $z = 0.15, 0.38, 0.51, 0.698, 1.48, \text{ and } 2.334$. The blue ellipsoids represent the regions constrained by the MTGP reconstructions, while the red markers or compact regions correspond to the Planck $\Lambda$CDM predictions. These visualizations serve as a powerful tool to provide a geometric perspective on the correlations and degeneracies between the parameters at each redshift. The ellipsoidal shapes encapsulate the relationships between the background and perturbation sectors, along with their mutual correlations. It also identifies potential deviations or tensions, with notable trends emerging at $z = 1.48$ and $z = 2.334$, where the reconstructed regions show significant departure from the Planck $\Lambda$CDM predictions. This suggests that these phase spaces not only validate the model at lower redshifts but also pinpoint redshift ranges where new physics or beyond-$\Lambda$CDM scenarios may need to be considered, which is consistent with our model-independent result in Fig. \ref{fig:rec_spheres}. 

The blue ellipsoids in Fig. \ref{fig:lcdm_spheres} depict parameter degeneracies and correlations at each redshift, derived from the reconstructed data. For instance, $H_0 r_d$ and $\Omega_m$ show strong correlations at lower redshifts ($z = 0.15, ~0.38, \text{ and } 0.51$), which gradually weaken with increasing redshifts at $z = 0.698$ and $z = 2.334$ respectively. Interestingly, at $z = 1.48$, the direction of the correlation notably shifts. The red ellipsoids correspond to the Planck $\Lambda$CDM best-fit predictions. The consistent overlap between the blue and red regions at lower redshifts affirms the $\Lambda$CDM model's validity. In contrast, the lack of overlap at higher $z$ presents challenges to explaining background and perturbation observables within the standard cosmological framework. This calls for scrutiny to discern whether it indicates potential departures from the concordance model arising from genuine physical phenomena or the influence of unaccounted systematic effects.

\begin{figure*}[t]
    \centering
    \includegraphics[width=\linewidth]{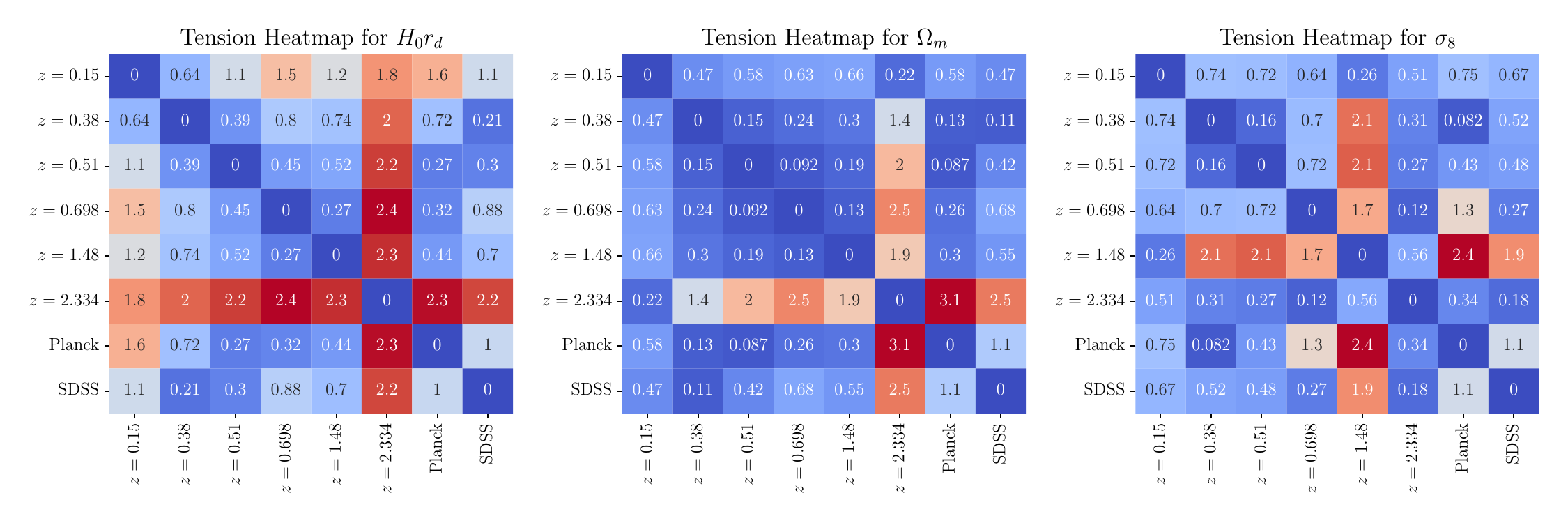}   
    \caption{Tension heatmaps between the $\Lambda$CDM model parameters $H_0 r_d$, $\Omega_m$, $\sigma_8$, across SDSS redshifts. We also show their comparison with the Planck and SDSS baseline results. }
    \label{fig:tension}
\end{figure*}

To better quantify the degree of statistical tension (measured in $\sigma$) in the $\Lambda$CDM model parameters $H_0 r_d$, $\Omega_m$, and $\sigma_8$ across different redshifts, we compute the Gaussian Tension. The heatmaps in Fig. \ref{fig:tension} reveal that tension becomes increasingly pronounced (exceeding $2\sigma$, highlighted in red) at higher redshifts, particularly at $z = 1.48$ and $z = 2.334$, where values diverge significantly from Planck and SDSS predictions. Conversely, regions in blue indicate tensions below $1\sigma$. At $z = 1.48$, $\sigma_8$ notably differs from those of lower redshifts and Planck/SDSS estimates. At $z = 2.334$, $H_0 r_d$ and $\Omega_m$ display $2\sigma$ tension with other redshifts and Planck/SDSS $\Lambda$CDM predictions. Finally, we summarize our findings as follows:

\squeezetable
\begin{table*}[t]
\setcounter{table}{2}
\caption{\label{tab:lcdm}%
Parameter Estimates for Reconstructed Cosmological Observables $D_M(z)/r_d$, $D_H(z)/r_d$ and $f\sigma_8(z)$ assuming $\Lambda$CDM model.
}
\begin{ruledtabular}
\begin{tabular}{l c c c c c c c c}
Sample & \multicolumn{2}{c}{${H_0}{r_d}$ [in 100 km/s]} & \multicolumn{2}{c}{$\Omega_{m}$} & \multicolumn{2}{c}{$\sigma_{8}$} & \multicolumn{2}{c}{$S_{8}$} \\
\cline{2-9}
& Best Fit & Mean with 1$\sigma$ & Best Fit  & Mean with 1$\sigma$ & Best Fit  & Mean with 1$\sigma$ & Best Fit  & Mean with 1$\sigma$ \\
\colrule
$z=0.15$ & $103.733 \pm 2.713$ & $104.128_{-3.190}^{+2.235}$ & $0.220 \pm 0.162$ & $0.184_{-0.120}^{+0.196}$ & $1.163 \pm 0.469$ & $1.003_{-0.265}^{+0.711}$  & $0.804 \pm 0.078$ &  $0.800_{-0.072}^{+0.077}$\\
$z=0.38$ & $101.226 \pm 2.828$ & $101.229_{-2.819}^{+2.785}$ & $0.306 \pm 0.076$ & $0.302_{-0.071}^{+0.078}$ & $0.816 \pm 0.057$ & $0.810_{-0.047}^{+0.057}$ & $0.812 \pm 0.082$ & $0.809_{-0.078}^{+0.083}$ \\
$z=0.51$ & $~99.776 \pm 2.429$ & $~99.760_{-2.412}^{+2.445}$ & $0.320 \pm 0.052$ & $0.318_{-0.049}^{+0.054}$ & $0.827 \pm 0.035$ & $0.826_{-0.035}^{+0.035}$ & $0.850 \pm 0.073$ & $ 0.848_{-0.071}^{+0.074}$ \\
$z=0.698$ & $~98.269 \pm 2.345$ & $~98.275_{-2.366}^{+2.343}$ & $0.326 \pm 0.040$ & $0.324_{-0.038}^{+0.041}$ & $0.864 \pm 0.039$ & $0.865_{-0.039}^{+0.039}$ & $0.900 \pm 0.074$ & $0.897_{-0.071}^{+0.076}$ \\
$z=1.48$ & $~96.626 \pm 5.500$ & $~96.735_{-5.622}^{+5.344}$ & $0.337 \pm 0.070$ & $0.328_{-0.059}^{+0.075}$ & $1.040 \pm 0.094$ & $1.037_{-0.088}^{+0.098}$ & $1.104 \pm 0.199$ & $1.083_{-0.173}^{+0.214}$ \\
$z=2.334$ & $122.055 \pm 9.804$ & $121.998_{-9.749}^{+9.768}$ & $0.183 \pm 0.042$ & $0.178_{-0.035}^{+0.045}$ & $0.894 \pm 0.246$ & $0.896_{-0.245}^{+0.239}$ & $0.695 \pm 0.208$ & $0.684_{-0.190}^{+0.210}$ \\ \hline 
Planck $\Lambda$CDM & $~99.078 \pm 0.925$ & $~99.076_{-0.918}^{+0.924}$ & $0.315 \pm 0.007$ & $0.315_{-0.007}^{+0.007}$ & $0.811 \pm 0.006$ & $0.811_{-0.006}^{+0.006}$ & $0.832 \pm 0.013$ & $0.832_{-0.013}^{+0.013}$ \\
SDSS $\Lambda$CDM & $100.589 \pm 1.204$ & $100.593_{-1.221}^{+1.188}$ & $0.297 \pm 0.015$ & $0.297_{-0.015}^{+0.016}$ & $0.850 \pm 0.035$ & $0.849_{-0.034}^{+0.036}$ & $0.846 \pm 0.042$ & $0.845_{-0.041}^{+0.042}$\\
\end{tabular}
\end{ruledtabular}
\end{table*}

\begin{itemize}
\item The parameter $H_0 r_d$ (in units of 100 km/s) shows a consistent decrease with redshift from $z=0.15$ to $z=1.48$, followed by a sharp increase at $z=2.334$, which emerges as an anomaly. Assuming the sound horizon at the drag epoch is constant at $r_d = 147.09 \pm 0.26$ Mpc, as inferred from early-Universe observations by Planck, the corresponding inferred values of $H_0$ mirror the trend in $H_0 r_d$. Specifically, $H_0$ is $70.792^{+1.519}_{-2.169}$ km Mpc$^{-1}$ s$^{-1}$ at $z=0.15$, $68.821^{+1.893}_{-1.916}$ km Mpc$^{-1}$ s$^{-1}$ at $z=0.38$, $67.822^{+1.662}_{-1.640}$ km Mpc$^{-1}$ s$^{-1}$ at $z=0.51$, $66.813^{+1.593}_{-1.609}$ km Mpc$^{-1}$ s$^{-1}$ at $z=0.698$, and $65.766^{+3.633}_{-3.822}$ km Mpc$^{-1}$ s$^{-1}$ at $z=1.48$. These values are consistent within $2\sigma$ with both the Planck 2018 estimate and the SH0ES 2021 measurement. In contrast, at $z=2.334$, $H_0$ shows a sharp rise to $82.941^{+6.641}_{-6.628}$ km Mpc$^{-1}$ s$^{-1}$, standing out as a significant outlier. 
\item The constraints on $\Omega_m$ agree with Planck $\Lambda$CDM predictions at redshifts $z=0.38$, $0.51$, and $0.698$ within the $1\sigma$ confidence level. While the best-fit values of $\Omega_m$ exhibit a gradually increasing trend, this variation is less pronounced compared to the trend observed in $H_0$. At $z=0.15$, the precision of $\Omega_m$ constraints is notably reduced due to the absence of BAO measurements for $D_M(z)/r_d$ and $D_H/r_d$ at this redshift, with the only available information coming from the RSD $f\sigma_8$ measurement from MGS tracers. Additionally, at $z=2.334$, $\Omega_m$ is significantly lower compared to the Planck baseline, breaking the increasing trend. This deviation, coupled with the anomalous sharp increase in $H_0 r_d$ (hence $H_0$), may indicate systematic effects or unexpected physics at this redshift bin. However, this result should be interpreted with caution, as there is no $f\sigma_8$ measurement from Ly-$\alpha$ tracers to robustly support this finding.
\item The $\sigma_8$ constraints remain relatively stable at redshifts $z=0.38$, $0.51$, and $0.698$, showing good agreement with the Planck baseline estimates within $2\sigma$. At $z=0.15$, the constraints exhibit significant broadening, indicating reduced precision, which can be attributed to the corresponding broadening of $\Omega_m$ in this redshift bin. At $z=1.48$, $\sigma_8$ shows a marked increase, resulting in a tension exceeding $3\sigma$ compared to Planck $\Lambda$CDM. This sharp rise at $z=1.48$ could be indicative of some form of rapid transition or deviation from standard cosmological expectations. Conversely, at $z=2.334$, $\sigma_8$ appears to decrease, but the large associated uncertainties render this result inconclusive.
\item The best-fit values of $S_8$ exhibit a gradual increasing trend with redshift from $z = 0.15$ up to $z = 1.48$. Within $1\sigma$, this pattern mirrors that of $\sigma_8$, demonstrating consistency with Planck $\Lambda$CDM predictions at lower redshifts $z = 0.15$, $0.38$, $0.51$, and $0.698$, followed by a statistically significant 2$\sigma$ rise at $z = 1.48$ that indicates a potential tension with Planck $\Lambda$CDM, pointing toward new physics affecting the perturbation sector. At $z = 2.334$, $S_8$ shows a subsequent decrease; however, the large uncertainties at this redshift preclude drawing definitive conclusions.
\end{itemize}

\subsection{Comparison with Complementary Datasets}

\begin{figure*}[t]
    \centering
     \includegraphics[width=\linewidth]{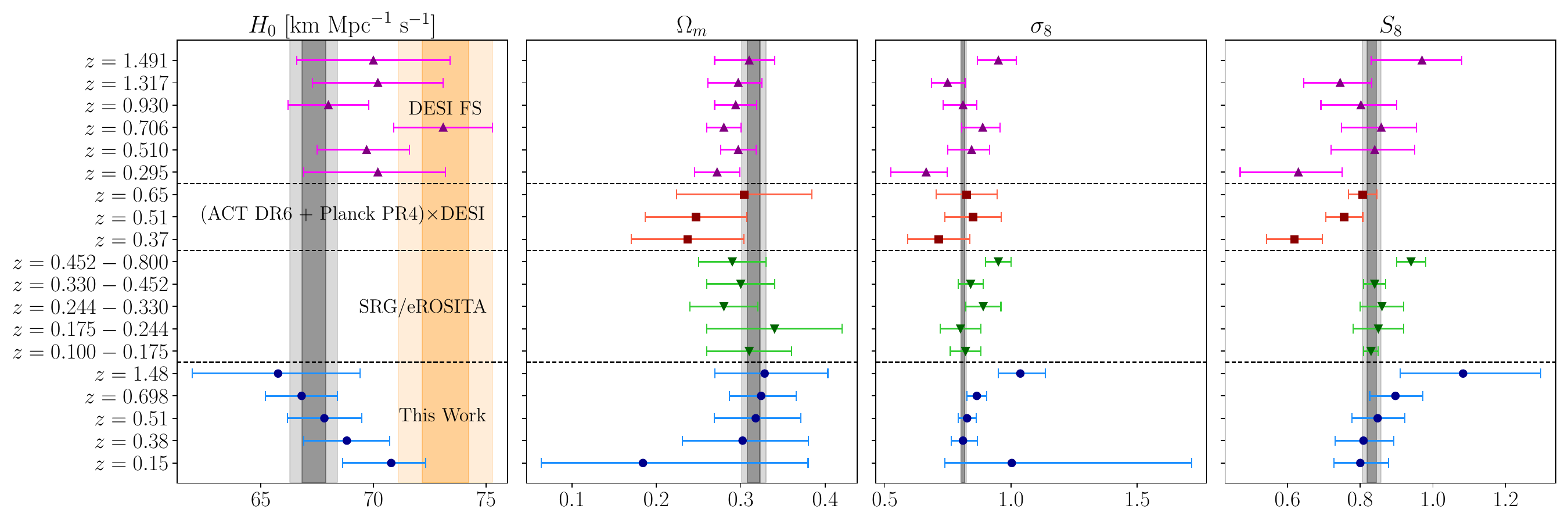}   
    \caption{Whisker plot showing the redshift-dependence on the inference of cosmological parameters, $H_0$, $\Omega_m$, $\sigma_8$, and $S_8$ obtained from fitting the $\Lambda$CDM model to the reconstructed functions. Comparison of these trends across multiple surveys.}
    \label{fig:whisker}
\end{figure*}

\begin{figure*}
    \centering
     \includegraphics[width=\linewidth]{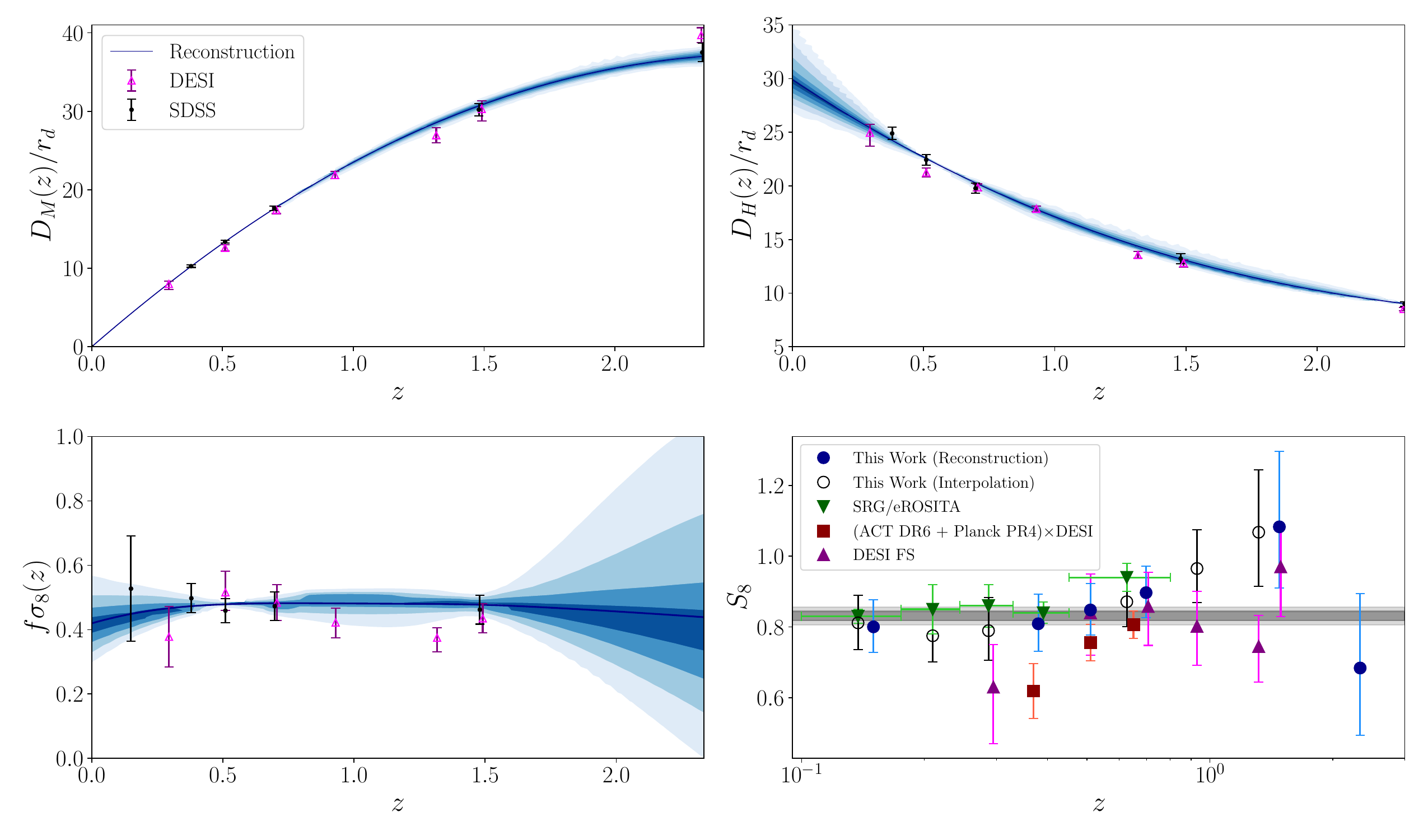}   
    \caption{Plots showing a comparison among the reconstructed SDSS observables, observational data from SDSS BAO+RSD as well as DESI BAO+FS surveys. The top-left panel presents reconstruction of $D_M(z)/r_d$ vs $z$, top-right panel shows $D_H(z)/r_d$ vs $z$, and bottom right panel shows $f\sigma_8(z)$ vs $z$. The bottom-right panel shows $S_8$ values from MTGP reconstruction of SDSS BAO+RSD data, compared to SRG/eROSITA, ACT DR6+Planck PR4+DESI BAO, and DESI FS constraints assuming $\Lambda$CDM.}
    \label{fig:S8}
\end{figure*}

The whisker plot in Fig. \ref{fig:whisker} offers a comparative visualization of constraints on the $\Lambda$CDM model parameters: $H_0$ (in units of km Mpc$^{-1}$ s$^{-1}$), $\Omega_m$, $\sigma_8$, and $S_8$, derived from various state-of-the-art surveys in the overlapping redshift range $z<1.5$. These include the results of our reconstruction at $z=0.15, ~0.38,~ 0.51, ~0.698, \text{ and } 1.48$ (referred to as ``This Work''), constraints based on the SRG/eROSITA catalogues \citep{Artis:2024zag}, the combination of ACT DR6 lensing + Planck PR4 + DESI BAO \citep{ACT:2024nrz}, and the DESI full-shape \citep{DESI:2024jis} analysis. The 1$\sigma$ and 2$\sigma$ regions for all parameters based on the Planck baseline are illustrated in gray shades, while the $H_0$ value from the SH0ES Collaboration is emphasized in orange. Each entry along the y-axis corresponds to a redshift bin associated with a specific dataset, while the horizontal error bars represent the uncertainty ranges for the respective parameters. Results obtained from our reconstruction are marked with blue circular markers, SRG/eROSITA with green downward triangles, (ACT DR6 + Planck PR4) $\times$ DESI BAO with red squares, and DESI FS with purple upward triangles. The visualization highlights the ability of our cosmological model-agnostic MTGP framework to provide precise and competitive constraints on $\Lambda$CDM parameters. Each panel focuses on a specific cosmological parameter, illustrating different facets of the derived constraints. Herein, we notice the following trends: \begin{itemize}
    \item Our reconstruction results reveal a consistent trend of $H_0$ increasing with decreasing $z$ in the range $0 < z < 1.48$: At $z = 0.698$, $H_0$ closely aligns with the Planck value, while at lower redshifts, such as $z = 0.15$ and $z = 0.38$, it gets closer to the SH0ES value or lies midway between SH0ES and Planck. 
    \item The DESI FS data exhibit an oscillating behavior in $H_0$, with the mean values alternatively increasing and decreasing between $z = 0.51$, $z = 0.706$, and $z = 0.930$. Specifically, for DESI FS, $H_0$ reaches the SH0ES value at $z = 0.706$ but reverts to the Planck value at $z = 0.930$. At the remaining redshifts, $H_0$ lies midway between the SH0ES and Planck results, stabilizing near $\approx$ 70 km Mpc$^{-1}$ s$^{-1}$.  
    \item For $\Omega_m$, our reconstruction results indicate that the central values of $\Omega_m$ increase with increasing values of effective redshift. However, a constant $\Omega_m$ remains consistent within 1$\sigma$ across the redshift range $z < 1.48$. 
    \item This increasing trend in $\Omega_m$ with $z_{\rm eff}$ is also observed in the results from (ACT DR6 + Planck PR4) $\times$ DESI BAO and DESI FS, although a constant $\Omega_m$ is permitted within 2$\sigma$. However, results from SRG/eROSITA show a constant $\Omega_m$ allowed within 1$\sigma$.
    \item For $\sigma_8$, our reconstruction results show that $\sigma_8$ consistently increasing with effective redshift. Results from SRG/eROSITA and (ACT DR6 + Planck PR4) $\times$ DESI also agree with this increasing trend in $\sigma_8$ vs $z_{\rm eff}$. With DESI FS, $\sigma_8$ follows a similar increasing trend up to $z = 0.7$, aligning with the trends observed in SDSS reconstruction up to $z = 0.698$. However, beyond this, $\sigma_8$ exhibits an anomalous decrease at $z = 0.93$ and $z = 1.317$. Since SDSS observations are not available in this redshift range, so definitive claims or comparisons cannot be made in this regard. 
    \item We notice a sharp increase in $\sigma_8$ at $z = 1.491$ from DESI FS, which is also obtained from our SDSS reconstruction. Interestingly, this deviation from Planck $\Lambda$CDM seen at $z = 1.48$ in SDSS and at $z = 1.491$ in DESI FS exceeds the $2.4\sigma$ and $1.8\sigma$ statistical limit, respectively. This feature, appearing in multiple generations of BAO and RSD data, may hint towards the possibility of new physics. 
    \item For SRG/eROSITA, a $3.2\sigma$ deviation from Planck $\Lambda$CDM is observed in $\sigma_8$ at $z = 0.452-0.800$ range. This contrasts with the $\sigma_8$ value at $z = 0.65$ from (ACT DR6 + Planck PR4) $\times$ DESI BAO, which remains consistent with the Planck $\Lambda$CDM prediction. For SDSS and DESI FS, at $z = 0.698$ and $z = 0.706$, respectively, $\Lambda$CDM is just included within the 1$\sigma$ confidence level.
    \item For $ S_8 $, our reconstruction shows an increase with $z_{\rm eff}$, a trend also observed for (ACT DR6 + Planck PR4) $\times$ DESI BAO. This trend is also visible in SRG/eROSITA results, where $ S_8 $ remains fairly constant up to $ z < 0.452 $, after which it increases strikingly in the range $ 0.425 < z < 0.825 $. 
    \item For low redshifts, $S_8 $ is $1.2\sigma$ and $2.48\sigma$ lower than the baseline Planck estimate at $ z = 0.295 $ for DESI FS and $ z = 0.37 $ for (ACT DR6 + Planck PR4) $\times$ DESI BAO respectively. However, our reconstruction with SDSS and constraints from SRG/eROSITA show that the low-$ z $ measurements align with the Planck estimate at 1$\sigma$. 
    \item For DESI FS, $ S_8 $ initially increases with $ z_{\text{eff}} $ up to $ z = 0.706 $, then at $ z = 0.93 $ and $ z = 1.317 $, it reverses direction and decreases as $ z $ increases. 
    \item At higher redshifts, however, we observe a notable increase in $S_8$ at $ z = 1.48 $ from our reconstruction as well as at $z=1.491$ from DESI FS, showing tension with Planck, with a deviation greater than 1$\sigma$.
\end{itemize}

Fig. \ref{fig:S8} showcases the reconstructed key cosmological observables in combination with SDSS BAO+RSD and DESI BAO+FS \citep{DESI:2024jis} data. The top-left panel presents the reconstruction of $D_M(z)/r_d$ as a function of redshift $z$. The top-right panel shows the reconstruction of $D_H(z)/r_d$ vs redshift $z$. The bottom-left panel displays reconstructed $f\sigma_8(z)$ over the same redshift range. The dark blue line represents the best-fit reconstruction from SDSS data, while the shaded regions denote the 1$\sigma$ and 2$\sigma$ confidence levels. The circle error bars in black correspond to SDSS data points, and the pink triangles represent data extracted from the DESI BAO+FS analysis. We undertake this comparison to understand the generic trends between the previous SDSS and the latest DESI datasets, exploring the implications of these trends in SDSS and follow up to investigate hints towards potential future trends in DESI. This includes examining how small differences between SDSS and DESI might affect the reconstruction of the functions, as a diagnostic check for the $\Lambda$CDM model across redshifts in a data-driven manner.
\begin{itemize}
    \item For $D_M/r_d$, the DESI low-$z$ measurements are consistent with SDSS data up to $z = 1.491$, where both values almost overlap, except for a minor dip observed at DESI QSO redshift $z = 1.317$ from the final reconstruction curve obtained using SDSS data. However, at higher redshifts, a >2$\sigma$ difference is observed in the case of the Ly-$\alpha$ tracer. This suggests that the reconstructed $D_M/r_d$ from DESI data is expected to deviate from that of SDSS beyond $z > 1.5$, as the Ly-$\alpha$ data at $z = 2.33$ will influence the training of MTGP hyperparameters. Consequently, this will lead to notable changes in the predicted values. The inferred $D_M/r_d$ from DESI may be comparatively higher compared to those of SDSS for redshifts $z>1.5$.
    \item The generic trend of $D_H/r_d$ for both SDSS and DESI remains quite similar throughout the redshift range $0 < z < 2.334$. However, slight differences are observed between the two datasets. For DESI, the LRG1 tracer at $z = 0.51$ yields a $D_H/r_d$ value that is lower than the corresponding SDSS value by more than 1$\sigma$. Similarly, at the DESI QSO tracer redshift $z = 1.317$, $D_H/r_d$ is found to be lower compared to the value obtained from the SDSS reconstruction. This suggests that future $D_H/r_d$ measurements from DESI could result in a steeper slope of the curve at these redshifts compared to that of SDSS. Such deviations may hint at new features to investigate as observational data continue to become more refined.
    \item The $f\sigma_8$ plot reveals interesting features: DESI exhibits an oscillatory behavior that is absent in SDSS, likely due to the fewer data points in the SDSS dataset. Nevertheless, both datasets remain consistent within 1$\sigma$. The presence of such oscillatory behavior in DESI can lead to more pronounced wiggles in the reconstructed function compared to the smoother reconstruction derived from SDSS. Additionally, the DESI BGS tracer at $z = 0.295$ shows a dip to lower values, suggesting that the reconstructed function will exhibit a larger dip at lower values compared to the current reconstruction based on SDSS MGS tracer data at $z = 0.15$.
\end{itemize}
The bottom-right panel of Fig. \ref{fig:S8} presents the $S_8$ values derived from MTGP reconstruction applied to SDSS data, with a LambdaCDM model fitted to the predicted constraints at each redshift bin. The panel also includes comparisons to constraints from SRG/eROSITA, ACT DR6 + Planck PR4 combined with DESI BAO, and DESI FS modeling, all assuming a LambdaCDM framework. The label 'This Work (Reconstruction)' represents the direct output of our reconstruction method at the SDSS effective redshifts, while 'This Work (Interpolation)' includes additional interpolated points to demonstrate the observed trends in $S_8$, highlighting its increase with effective redshift. Resulting constraints from SRG/eROSITA, ACT DR6 + Planck PR4 + DESI, and DESI FS are shown, emphasizing the competitive constraints provided by our methodology. To better capture the features at low-$z$, the x-axis is scaled logarithmically. A general increasing trend in $S_8$ is observed across all datasets. However, the value at $z = 2.334$, corresponding to the Ly-$\alpha$ measurement from the SDSS reconstruction, appears to be an outlier. Notably, there is no $f\sigma_8$ measurement at this redshift, which limits its statistical significance.

\section{Conclusion} \label{sec:conclusion}

The $\Lambda$CDM model, while phenomenologically successful in describing the dynamics of our Universe, remains a parameterized framework with no underlying theoretical explanation for its core components, such as dark energy and dark matter. However, tensions in model parameters—such as the $>5\sigma$ discrepancy in $H_0$ between local and CMB measurements, and the $\sim2-2.5\sigma$ mismatch in $S_8$ between CMB and weak lensing surveys—raise questions about its validity. These tensions, coupled with emerging evidence for redshift-dependence in $H_0$, $\Omega_m$, and $S_8$, suggest that either modifications to $\Lambda$CDM, its underlying assumptions are required, or unaccounted systematic effects when combining datasets must be addressed.  

Traditional methods of stress-testing $\Lambda$CDM consistency across redshifts often rely on binning mechanisms, which inherently lose resolution and fail to capture subtle trends or correlations. In this work, we present a novel approach using the MTGP framework to reconstruct cosmological observables across redshifts, simultaneously probing both background ($D_M/r_d$, $D_H/r_d$) and perturbation ($f\sigma_8$) sectors. Our analysis is based solely on SDSS-IV eBOSS data, incorporating the full covariance of the dataset, which includes auto-correlations of the same cosmological function and cross-correlations between different functions at various effective redshifts. By accounting for all systematics within the dataset and refraining from combining data from multiple surveys, we mitigate the influence of inter-survey systematics that could compromise the robustness of our results. This approach also avoids potential confirmation bias that can arise when datasets are combined under specific model assumptions, ensuring an unbiased evaluation of $\Lambda$CDM. 

In this work, we performed an MTGP reconstruction of the SDSS-IV eBOSS BAO and RSD observables, enabling the construction of phase space volumes at each of the SDSS effective redshifts. This analysis utilized the full correlated SDSS BAO+RSD dataset, incorporating auto- and cross-correlations between observables to capture their full covariance structure. Within these reconstructed volumes, we evaluated constraints on cosmological parameters—$H_0 r_d$, $\Omega_m$, $\sigma_8$, and $S_8$ - under the assumption of Planck $\Lambda$CDM as the underlying model. By adopting $r_d$ derived from early-universe physics as determined by Planck, we derived $H_0$ values at each binned redshift and analyzed the redshift-dependent behavior of these parameters. The trends were examined to assess deviations from $\Lambda$CDM predictions, and quantify the degree of tension with the baseline Planck.  

Our results showed that at low redshifts ($z < 0.7$), the reconstructed observables are in agreement with $\Lambda$CDM predictions. However, at $z = 1.48$, we identify deviations in the reconstructed $f\sigma_8$ values compared to the Planck $\Lambda$CDM model. At $z = 2.334$, the reconstructed $D_M/r_d$ and $D_H/r_d$ observables exhibit significant deviations from $\Lambda$CDM predictions. Furthermore, we found redshift-dependent trends in the model parameters when the reconstructed volumes are fit to the $\Lambda$CDM model. For instance, $H_0$ decreases with increasing effective redshift, while $\Omega_m$ shows an increasing trend, although a constant $\Omega_m$ remains consistent within $1\sigma$ constraints. Both $\sigma_8$ and $S_8$ exhibit an increase with rising $z_\text{eff}$, with a sharp and striking increase observed at $z = 1.48$. These trends highlight the need for extensions to the standard cosmological model or a better understanding of systematic uncertainties that merit further investigation.

The trends we observed in $H_0$, $\Omega_m$, $\sigma_8$, and $S_8$ both corroborate and challenge findings from those in the existing literature. The observed redshift-dependent variations in $H_0$ are consistent with  \citet{H0LiCOW:2019pvv, Krishnan:2020obg, Millon:2019slk, Krishnan:2020vaf, Dainotti:2021pqg, Colgain:2022nlb, Colgain:2022rxy, Hu:2022kes, Jia:2022ycc, Vagnozzi:2023nrq}, supporting the notion of decreasing $H_0$ with increasing $z$. For $\Omega_m$, we find an increase with $z$, which aligns with the trends observed by \citet{Colgain:2024mtg, Colgain:2022nlb, Colgain:2022rxy, Risaliti:2018reu, Lusso:2020pdb, Yang:2019vgk, Khadka:2020vlh, Pasten:2023rpc}, although being compatible with studies like \citet{Dinda:2023xqx, Artis:2024zag, DESI:2024jis, DESI:2024hhd, DESI:2024mwx} at $1\sigma$ that suggest no such evolution. Similarly, the parameters $\sigma_8$ and/or $S_8$ show evidence of evolution with redshift, agreeing with findings from \citet{Adil:2023jtu, Akarsu:2024hsu, ACT:2024nrz, DESI:2024jis, Artis:2024zag}, indicating that the amplitude of matter fluctuations changes over cosmic time, but contrasting with results from \citet{Poulin:2022sgp, Manna:2024wak, DES:2021wwk}, which find no significant redshift-dependent variations in $\sigma_8$ or $S_8$.

The MTGP framework offers several key advantages, making it a powerful tool for cosmological analysis. It is inherently model-independent, avoiding assumptions tied to specific cosmological models and enabling unbiased diagnostic tests. Integrating correlations between background and perturbation observables provides a unified and holistic view of cosmological trends. Unlike traditional binning approaches, which often obscure subtle variations, the MTGP framework captures smooth, high-resolution trends across redshifts, revealing potential inconsistencies that might otherwise go unnoticed. The phase spaces generated through this approach further enhance its utility by visualizing the overlap between background and perturbation sectors, allowing for a detailed examination of inconsistencies and the evolution of features across redshifts. 

This method will be especially relevant for ongoing surveys like DESI and can be directly applied once the DESI data vector, along with the covariance matrix from full-shape modelling of galaxy clusters \citep{DESI:2024jis}, is publicly released. These results, along with DESI DR1 BAO analysis, indicate a preference for dynamical dark energy over $\Lambda$CDM, with $\Lambda$CDM being excluded at more than $2\sigma$ in the Planck+DESI+Pantheon+, Planck+DESI+DES-SN5YR and Planck+DESI+Union3 analyses \citep{DESI:2024hhd, DESI:2024mwx}. Additionally, upcoming surveys like Euclid \citep{Euclid:2019clj} can provide us with separate measurements of $f(z)$, $\sigma_8(z)$, and $f\sigma_8(z)$, by combining RSD measurements in the power spectrum and bispectrum \citep{Gil-Marin:2016wya} or with galaxy-galaxy lensing data \citep{delaTorre:2016rxm, Shi:2017qpr, Jullo:2019lgq}, thereby breaking the inherent degeneracy. So, the ability of our framework to simultaneously analyze both background and perturbation sectors will henceforth be crucial. 

It is crucial to recognize the assumption of a fiducial cosmology when measuring the BAO signal from galaxy surveys. This parameterized template is essential for converting redshifts into distances and defining the input parameters for the BAO reconstruction algorithm, which introduces an inherent model dependence in the BAO data extraction. Although the final analysis allows for deviations from the assumed cosmology, the reliance on a specific model during the initial stages represents a form of data compression that can bias the results toward the assumptions of the fiducial framework. For instance, the BAO analysis of the SDSS-IV eBOSS data employs a fiducial cosmology based on Planck $\Lambda$CDM, which could subtly imprint its assumptions into the extracted distance measurements. 

While our study mitigates systematic uncertainties by refraining from combining datasets across surveys, the reliance on fiducial cosmology in the SDSS BAO analysis could still affect the robustness of reconstructed observables. Such fiducial cosmology-dependent systematics has been evaluated in recent work on DESI 2024 BAO analysis, where the impact of varying the fiducial cosmology was tested using mock catalogues spanning alternative cosmological scenarios, including a lower cold dark matter density, dynamical dark energy, and changes in the amplitude of matter clustering \citep{DESI:2024ude} demonstrating that fiducial-cosmology-dependent systematics contribute a small but non-negligible error, estimated at 0.1\% for isotropic and anisotropic parameters. This underscores the importance of future analyses, such as those from DESI and Euclid, which are designed to minimize model assumptions during data extraction, enabling more robust, model-independent reconstructions of cosmological trends.

Finally, the growing importance of redshift-dependent studies in cosmology highlights the need for tools that can uncover subtle deviations from $\Lambda$CDM. MTGP-based reconstruction sets the stage for future investigations, paving the way for exploring new physics while maintaining robustness against systematic uncertainties inherent in multi-survey combinations.

\begin{acknowledgments}
We are grateful to Shadab Alam, Eoin \'{O} Colg\'{a}in, Shahin Sheikh Jabbari and \"{O}zg\"{u}r Akarsu for useful discussions. PM acknowledges funding from the Anusandhan National Research Foundation (ANRF), Govt of India under the National Post-Doctoral Fellowship (File no. PDF/2023/001986). AAS acknowledges the funding from ANRF, Govt of India under the research grant no. CRG/2023/003984. We acknowledge the use of HPC facility, Pegasus, at IUCAA, Pune, India. This article/publication is based upon work from COST Action CA21136 – “Addressing observational tensions in cosmology with systematics and fundamental physics (CosmoVerse)”, supported by COST (European Cooperation in Science and Technology).
\end{acknowledgments}

%

\vspace{5mm}


\software{numpy \citep{2020NumPy-Array},
          scipy \citep{2020SciPy-NMeth}, 
          matplotlib \citep{Hunter:2007},
          jax \citep{jax2018github}
          tinygp \citep{tinygp}, 
          numpyro \citep{numpyro, bingham2019pyro}, 
          cobaya \citep{Torrado:2020dgo},
          GetDist \citep{Lewis:2019xzd},
          fgivenx \citep{fgivenx}
          }






\bibliography{references}{}
\bibliographystyle{aasjournal}



\end{document}